\documentclass[12pt]{iopart}
\usepackage{graphics}
\usepackage{amssymb}
\usepackage{citesort}

\newcommand{\ie}{{\it i.e.\ }}
\newcommand{\eg}{{\it e.g.\ }}

\newcommand{\kgsm}{$\mathrm{kg\,m^{-2}}$}
\newcommand{\kgm}{$\mathrm{kg\,m^{-1}}$}

\newcommand{\be}{\begin{equation}}
\newcommand{\bea}{\begin{eqnarray}}
\newcommand{\ee}{\end{equation}}
\newcommand{\eea}{\end{eqnarray}}
\renewcommand{\etal}{{\it et al.}}

\begin{document}

\title[Spatial variability of fibrous materials]{Spatial variability of void structure in thin stochastic fibrous materials}
\date{\today}

\author{W.W.\ Sampson}
\address{School of Materials, University of Manchester, Manchester, M13 9PL, UK}
\ead{w.sampson@manchester.ac.uk}


\begin{abstract}
Theory is presented for the distributions of local process intensity and local average pore dimensions in random fibrous materials. For complete partitioning of the network into contiguous square zones, the variance of local process intensity is shown to be proportional to the mean process intensity and inversely proportional to the zone size. The coefficient of variation of local average pore area is shown to be approximately double that of the local average pore diameter with both properties being inversely proportional to the square root of zone size and mean process intensity. The results have relevance to heterogenous near-planar fibrous materials including paper, nonwoven textiles, nanofibrous composites and electrospun polymer fibre networks.

\end{abstract}

\pacs{61.43.Gt; 61.43.Bn; 61.43.-w; 81.05.Qk; 81.05.Rm}

\maketitle
\section{Introduction}
The global average pore size of thin, \ie near-planar, heterogeneous fibrous materials and its distribution have been widely studied using statistical geometry and simulation. Typically the context of these studies has been materials with widespread application in society and industry such as paper~\cite{cl,ds96}, nonwoven textiles~\cite{ad85}, and fibrous filter media~\cite{pk,ostoja}. Interest in the structural characteristics of fibrous materials has increased in recent years as researchers seek to develop materials for future applications such as carbon nanotube `buckypaper'~\cite{endo,hall}, nanofibrous composites~\cite{paul,eichhornreview} and electrospun fibrous networks for application as scaffolds in tissue engineering~\cite{pham,sill}.

\begin{figure}
\begin{flushright}
{\resizebox{13 cm}{!}{\includegraphics{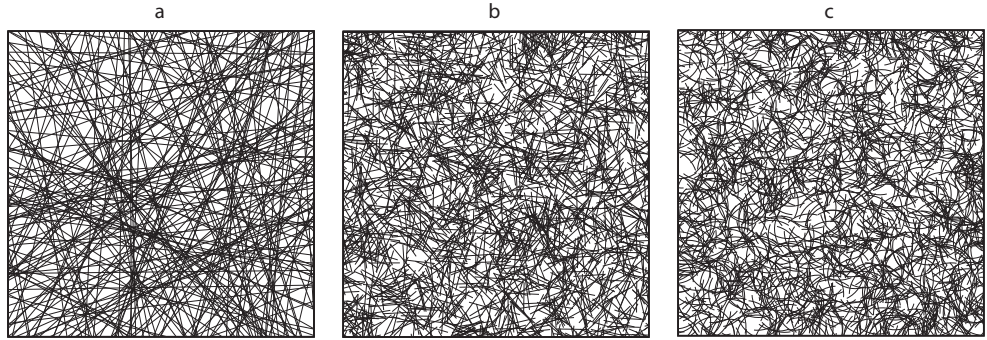}}}
\caption{Graphical representations of Poisson line processes in a unit square with intensity~$\bar\tau = 200$. a) infinite lines; b) straight lines with length~0.1; c) curved lines with length and radius of curvature~0.1.} \label{F:3Nets}
\end{flushright}
\end{figure}

In a seminal paper, Miles~\cite{miles64} provides several properties of the polygons generated by the stochastic division of a plane by a Poisson process of straight lines with infinite length. A graphical representation of such a process is given in Figure~\ref{F:3Nets}a. The process intensity is characterized by the expected line length per unit area,~$\bar\tau$. The following results of Miles are utilized in the present study:
\begin{quote}
\begin{itemize}
\item The expected area of polygons is
\be\label{e:abarMiles}
\bar{a} = \frac{\pi}{{\bar\tau}^2}\ .
\ee
\item The distribution of diameters of the largest circle inscribed within polygons is exponential with mean
\be\label{e:dbarMiles}
\bar{d} = \frac{1}{\bar\tau}
\ee
\item $\bar{a}$ and~$\bar{d}$ are independent of the width of lines, for any probability density of line widths.
\end{itemize}
\end{quote}
Miles showed also that the expected number of sides per polygon is~4 and that the fraction of triangles is~$P(3) = (2 - \frac{\pi^2}{6}) \approx 0.355$; Tanner~\cite{tanner} obtained the fraction of quadrilaterals, $P(4)\approx 0.381$. The fractions of polygons with more than~4 sides are not known analytically, but have been obtained by Monte Carlo methods~\cite{crainmiles,george}. We note that very similar results were obtained by Piekaar and Clarenburg~\cite{pk} from simulations of networks of fibres with finite length. Crain and Miles~\cite{crainmiles} observed the probability of $n$-sided polygons was well approximated by the Poisson variable $(n-3)$ with mean $(\bar{n} -3) = 1$. This result was used by Dodson and Sampson~\cite{PlinProp} to approximate the probability density functions for polygon areas and perimeters, assuming regular polygons. The coefficient of variation of polygon area was close to~2 and that of polygon perimeter was~$\sqrt{3}/2$; despite the assumption of regularity, these are close to the analytic results of Miles~\cite{miles64}.

Now, each intersection between lines represents one of the vertices of four polygons and for a Poisson process of infinite lines, Miles~\cite{miles64} gives the expected number of intersections per unit area as~${\bar\tau}^2/\pi$, \ie $1/\bar{a}$. For the case of networks lines with finite length and constant width (Figure~\ref{F:3Nets}b), the expected number of intersections per unit area was derived by Kallmes and Corte~\cite{kalcort1} for infinite networks; in practical contexts, this can be considered satisfied of samples are large relative to the dimensions of voids. Recently it has been shown~\cite{EichSamp2} that Kallmes and Corte's expression for finite lines is the same as that of Miles for infinite lines. Further, Berhan~\etal~\cite{berhan} showed the same dependency by considering the probability of intersection of
segments of finite length curved fibres (\eg Figure~\ref{F:3Nets}c); validity was confirmed via simulation studies for networks of
fibres with sinusoidal curvature with differing sinusoidal
frequency. Accordingly, we may be confident that theory describing a planar Poisson process of infinite length lines provides an appropriate reference to probe the void structure of networks of finite and curved fibres, greatly simplifying the analysis.

Statistical geometric models for the polygon area and pore diameter distribution suggest that these are described by gamma distributions~\cite{ds97,mppsd,chatterjee} and this has been confirmed through simulation~\cite{ostoja} and by measurements on paper~\cite{wwsAppita,JFBloch}, nonwoven textiles~\cite{jaganathan} and electrospun polymer nanofibre networks~\cite{tomba}. Whereas these models and measurements characterize the global average pore dimensions and its distribution, preliminary inspection of Figure~\ref{F:3Nets} reveals significant local differences in pore dimensions corresponding to the underlying variability in process intensity. Quantification of local variability in pore dimensions is the focus of this study.

The variance of local process intensity, and hence that of local mass density was derived for a planar Poisson process of rectangles by Dodson~\cite{ctjdJRSS} and has since been widely employed as the standard reference in quantifying variability in paper~\cite{deng}. Relationships between network uniformity and fracture behaviour have been demonstrated for this material~\cite{uesaka,wathen}. Similarly, flow and filtration efficiency in fibrous filters exhibit variability between regions~\cite{schweers,huang}. In a theoretical treatment, Chatterjee~\cite{chatterjee2} considered the local average elastic modulus of nanofibrous composites and demonstrated a strong sensitivity to the pore size of the constituent fibre network. Holzmeister~\etal\ varied local pore size in electrospun polymer scaffolds for tissue engineering by generating networks of fibres with bimodal distributions of diameter and remarked that the tendency of cells to proliferate on these scaffolds was influenced by local void dimensions~\cite{holzmeister}.

Here we consider the distribution of local process intensity for a Poisson process of infinite lines. From this we obtain approximate probability density functions for the distribution of local averages of pore area and pore diameter in planar stochastic fibrous materials.

\section{Variance of local process intensity}
We seek the variance of local process intensity for a planar Poisson line process of mean intensity,~$\bar\tau$, partitioned into contiguous square zones of side,~$x$. We denote the local process intensity within such zones,~$\widetilde\tau$.

Before proceeding, it is helpful to obtain estimates of the likely range of~$\bar\tau$ encountered in real materials. If the expected mass per unit area, or `areal density', of a fibre network is, $\bar\beta$~(\kgsm), and the linear density of the constituent fibres is~$\delta$~(\kgm), then the expected process intensity, $\bar\tau = \bar\beta/\delta$. For materials such as paper and nonwoven textiles, with mass per unit area between~20 and~100~$\mathrm{g\,m^{-2}}$ we expect~$\bar\tau$ to be of order~$10^2\,{\rm mm}^{-1}$; for electrospun networks of polymer fibres with mass per unit area around~10~$\mathrm{g\,m^{-2}}$  and fibre diameter around~1~$\mathrm{\mu m}$ we expect~$\bar\tau$ to be of order~$10^3\,{\rm mm}^{-1}$.

From Coleman~\cite{coleman} the probability density of the length,~$l_1$,  of random secants in a unit square is
\be\label{e:pdfcolemean}
f(l_1) = \left\{\begin{array}{ll}
\frac{2\,l_1}{\pi} &  \mbox{if $0 < l_1 \leq 1$}\\
\frac{4}{\pi\,l_1\,\sqrt{l_1^2 - 1}} - \frac{2\,l_1}{\pi} & \mbox{if $1< l_1 \leq \sqrt{2}$}\\
0 & \mbox{otherwise.}
\end{array} \right.
\ee
\noindent with mean $\bar{l}_1 = 4\,\left(1 - \sqrt{2} + 3\,\log(1 + \sqrt{2})\right)/(3\,\pi) \approx 0.946$.

The expected total length of lines in a square zone of side~$x$ is
\be
\bar{L} = \bar\tau\,x^2\ .
\ee
\noindent We denote the length of secants in a square of side~$x$, $l_x$, with expected length, $\bar{l}_x  = \bar{l}_1\, x$. It follows that the expected number of secants in a square of side~$x$ is
\be\label{e:Nbar}
\bar{N} = \frac{\bar{L}}{\overline{{l}_x}} = \frac{\bar\tau\,x}{\bar{l}_1} \ \ .
\ee

If the number of lines contained in a given square of side~$x$ is~$\widetilde{N}$, then the total line length in that square is
\be
\widetilde{L} = \sum_{i=0}^{\widetilde{N}} l_{x,i}\ \ ,
\ee
\noindent such that the local process intensity is
\be
\widetilde\tau = \frac{\widetilde{L}}{x^2}\ \ .
\ee
\noindent It follows that the variance of local process intensity is given by
\be
\sigma_x^2(\widetilde\tau) = \frac{\sigma_x^2(\widetilde{L})}{x^4}\ \ ,
\ee
\noindent where the subscript~$x$ is included to denote that the variance depends on the zone size. Note also that~${\sigma_x^2(\widetilde{L})} = x^2\,{\sigma^2(\widetilde{L}_1)}$, where~$\widetilde{L}_1$ is the total line length in a unit square containing~$\widetilde{N}$ secants. Thus,
\be\label{e:VarTau1}
\sigma_x^2(\widetilde\tau) = \frac{\sigma^2(\widetilde{L}_1)}{x^2}\ \ .
\ee
Now,~$\widetilde{L}_1$ is a random variable obtained as the sum of the lengths of~$\widetilde{N}$ independent and identically distributed secants in a unit square, \ie $\widetilde{L}_1 = \sum_{i=0}^{\widetilde{N}} l_{1,i}$, where $l_{1,i}$ is a continuous random variable with probability density given by Equation~(\ref{e:pdfcolemean}) and~$\widetilde{N}$ takes integer values. The mean and variance of~$\widetilde{L}_1$ are given by~\cite{CRC_SMTF}
\bea
\bar{L}_1 & = & \bar{N}\,\bar{l}_1 \\
\sigma^2(\widetilde{L}_1) & = & \bar{N}\,\sigma^2(l_1) + {\overline{l}_1}^2\, \sigma^2(\widetilde{N})\ \ .
\eea
We assume that~$\widetilde{N}$ is a Poisson random variable, such that $\sigma^2(\widetilde{N}) = \bar{N}$
and we have
\be
\sigma^2(\widetilde{L}_1) =  \bar{N}\left(\sigma^2(l_1) + {\overline{l}_1}^2\right)
\ee
\noindent Now, $\sigma^2(l_1) = \overline{l_1^2} - {\overline{l}_1}^2$. So
\be\label{e:VarL1a}
\sigma^2(\widetilde{L}_1) =  \bar{N}\,\overline{l_1^2}\ \ .
\ee
\noindent We obtain $\overline{l_1^2}$ as the second moment of the probability density given by Equation~(\ref{e:pdfcolemean}):
\be
\overline{l_1^2}  =  \int_0^{\sqrt{2}} l_1^2\,f(l_1)\, \rmd l_1 = \frac{3}{\pi}\ . \label{e:l1quaredbar}
\ee
\noindent For completeness, we note that $\sigma^2(l_1) = (3/\pi) - {\bar{l}_1}^2 \approx 0.0593$.

Substituting Equations~(\ref{e:VarL1a}), (\ref{e:l1quaredbar}) and~(\ref{e:Nbar}) in Equation~(\ref{e:VarTau1}) yields our final expression for the variance of local process intensity for square zones of side~$x$:
\bea
\sigma_x^2(\widetilde\tau) & = & \frac{3}{\pi}\,\frac{\overline{N}}{x^2} \label{e:VarTauN}\\
& = & \frac{3}{\pi\,\overline{l}_1}\,\frac{\bar\tau}{x} \nonumber\\
& \approx & \frac{\bar\tau}{x}\ .\label{e:bartauApprx}
\eea

\subsection{Distribution of local process intensity}

From the Central Limit Theorem, we expect the distribution
of local process intensity,~$\widetilde\tau$ to be well approximated by a Gaussian distribution if the expected number of secants in a square zone of side~$x$ is sufficiently large. For low intensity processes and at small~$x$ we anticipate that the distribution of local process intensity will exhibit a positive skew as a consequence of the underlying Poisson process for~$\widetilde{N}$.

Derivation of the distribution of~$\widetilde\tau$ has proved intractable, so here we estimate the skewness of the distribution by considering a Poisson process of secants in a unit square. The influence of changing~$\bar\tau$ or~$x$ is therefore captured entirely by varying the expected number of secants in the unit square,~$\bar{N}$. From Equations~(\ref{e:Nbar}) and~(\ref{e:VarTauN}), the mean and variance are $\bar\tau = \bar{N}\,\bar{l}_1$ and $\sigma^2(\widetilde\tau) = 3\,\bar{N}/\pi$. Noting that~$\bar{l}_1 \approx \overline{l_1^2} \approx 1$ and given that $0 \leq \l_1 \leq \sqrt{2}$, we assume that $\overline{l_1^3} \approx 1$ such that the third central moment,~$\mu_3$, of the probability density of~$\widetilde\tau$ is approximated by that of the underlying Poisson process, \ie $\mu_3(\widetilde\tau) \approx \mu_3(N) = \bar{N}$. The skewness of the distribution of~$\widetilde\tau$ is therefore approximated by
\bea
\gamma_1(\widetilde\tau) & = &  \frac{\mu_3(\widetilde\tau)}{\sigma^3(\widetilde\tau)}\\
               & \approx & \frac{\pi^\frac{3}{2}}{3\sqrt{3}\,\sqrt{\bar{N}}}\label{e:approxskew}
\eea
To test the validity of Equation~(\ref{e:approxskew}) and to probe the distribution of~$\widetilde\tau$ further, a Monte-Carlo simulation of random lines in a unit square was carried out using {\em Mathematica}~\cite{mma}. The number of lines in a unit square was generated as a Poisson random variable with mean~$\bar{N}$. For each random line, the code generated a pair of points drawn from a standard uniform distribution to represent a coordinate through which the line passed. This coordinate was associated with an angle drawn from a uniform distribution on the interval $\left[-\pi/2, \pi/2\right)$. The equation of the line was generated from the coordinate and the angle; its points of intersection with the perimeter of the unit square were calculated, allowing the length of the line to be computed. For each~$\bar{N}$, $10^6$ unit squares containing~$N$ lines were simulated and the local process intensity computed as the total length of the lines in each square. As a check, the distribution of~$10^6$ random lines was compared with the probability density given by Equation~(\ref{e:pdfcolemean}); the mean agreed within~0.003~\% and the variance agreed within~0.11~\%.

\begin{figure}
\begin{center}
{\resizebox{8 cm}{!}{\includegraphics{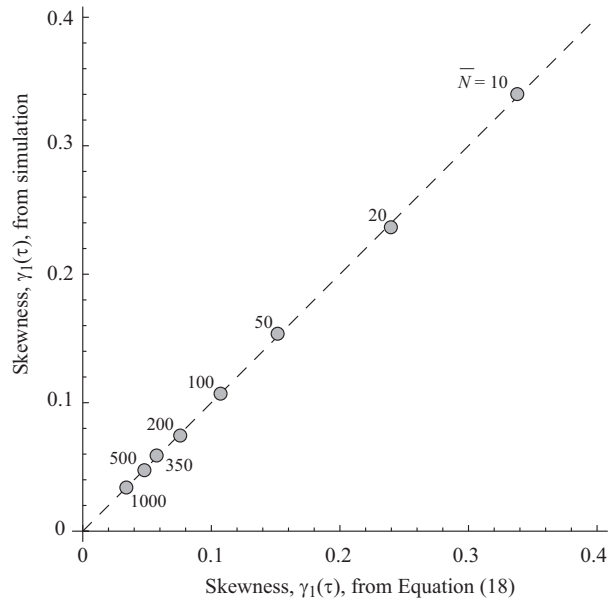}}}
\caption{Skewness of simulation data plotted against that calculated using Equation~(\protect\ref{e:approxskew}). Data labels give mean number of lines in a unit square; broken line has unit gradient.} \label{F:SkewTest}
\end{center}
\end{figure}

The skewness of the local process intensity,~$\gamma_1(\widetilde\tau)$, of the simulation data arising from different input~$\overline{N}$ is plotted against that obtained using Equation~(\ref{e:approxskew}) in Figure~\ref{F:SkewTest}. Data labels represent the value of~$\overline{N}$ input to the simulation and the broken line has unit gradient. A linear regression on the data has gradient~0.9994 with coefficient of determination,~$r^2 = 0.9997$.

\begin{figure}\begin{flushright}
{\resizebox{14 cm}{!}{\includegraphics{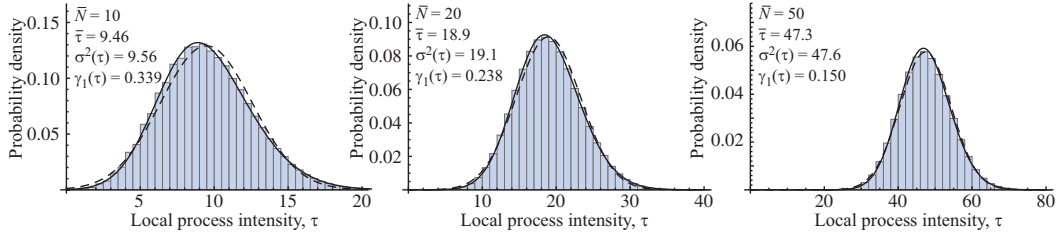}}}
\caption{Histograms showing the frequency of local process intensity,~$\tau$, arising from Monte-Carlo simulations of
random lines in a unit square. Heights of bars have been scaled to give unit area under the curve. Solid line represents PDF of
fitted skew-normal distribution; broken line represents PDF of Gaussian distribution
with same mean and variance as the data.} \label{F:HistoN10N20N50}
\end{flushright}
\end{figure}

Histograms of the local process intensity arising from the simulations are plotted in Figure~\ref{F:HistoN10N20N50} for~$\bar{N} = 10, 20$ and~50. To approximate the probability density of the data, the heights of the bars are given by the frequency divided by the bin width. The solid lines represent the probability densities of skew-normal distributions fitted by a least-squares method to the cumulative data. The probability density of the skew-normal distribution is~\cite{Azzalini}

\be\label{e:SNpdf}
g(\widetilde\tau) = \frac{e^{-\frac{(\widetilde\tau - m)^2}{2 s^2}} \,\rm{erfc}\left(\frac{\alpha\,(m - \widetilde\tau)}{\sqrt{2} s}\right)}{\sqrt{2\,\pi}\,s}\ \ ,
\ee

\noindent where $\rm{erfc}(\zeta)$ is the complementary error function. The mean, variance and skewness are given by
\bea
\bar\tau & = & m + \frac{\sqrt{2}\,\alpha\,s}{\sqrt{\pi}\,\sqrt{1+\alpha^2}} \nonumber \\
\sigma^2(\widetilde\tau) & = & \left(1 - \frac{2\,\alpha^2}{\pi\,(1+\alpha^2)}\right) s^2 \nonumber \\
\gamma_1(\widetilde\tau) & = & \frac{\sqrt{2}\,(4 - \pi)\,\alpha^3}{\left(\pi + (\pi - 2)\,\alpha^2\right)^\frac{3}{2}}\nonumber
\eea
\noindent respectively.

As anticipated from the skewness values plotted in Figure~\ref{F:SkewTest}, the distributions are increasingly well approximated by a Gaussian probability density as~$\bar{N}$ increases. When~$\bar{N}$ is greater than about~50, the skewness is negligible and the Gaussian can be assumed to describe the distribution of local process intensity well.

\subsection{Distribution of local average pore area}

We proceed assuming the distribution of local process intensity to be Gaussian with mean~$\bar\tau$ and variance~$\bar\tau/x$ ({\it cf.} Equation~(\ref{e:bartauApprx})), \ie
\be\label{e:GaussTau}
g(\widetilde\tau) = \sqrt{\frac{x}{2\,\pi\,\bar\tau}}\,e^{-\frac{x\,\left(\widetilde\tau - \bar\tau\right)^2}{2\,\bar\tau}}\ \ .
\ee
\noindent We note that for $\bar\tau\,x > 20$, $\int_{-\infty}^0 g(\widetilde\tau)\,\rmd \widetilde\tau < 4 \times 10^{-6}$; accordingly, truncation of the distribution such that $0 \leq \widetilde\tau < \infty$ is unnecessary for practical purposes.

From Equation~(\ref{e:abarMiles}), we expect the local average pore area to be
\be\label{e:atilde}
\widetilde{a} = \frac{\pi}{{\widetilde\tau}^2}\ \ .
\ee
Inevitably, when a network is partitioned into contiguous square zones, some polygons intersect the perimeter of the zone. The expected number of polygons intersecting the perimeter of a square zone of side~$x$ can be approximated as~$n_{\rm perim} = 4\,x/\bar{d} = 4\,x\,\bar\tau$ and the expected number of polygons in the square is~$n_{\rm area} = x^2/\bar{a} = x^2\,\bar\tau^2/\pi$. So, the expected fraction of polygons intersecting the perimeter of the square is approximately~$n_{\rm perim}/n_{\rm area} = 4\,\pi/(\bar\tau\,x)$. When~$\bar\tau = 100\,\rm{mm}^{-1}$ and~$x = 1~\rm{mm}$ this fraction is~$0.125$ and $n_{\rm area} > 3000$ so it is reasonable to assume that Equation~(\ref{e:atilde}) provides a good measure of the local average polygon area.

The probability density of local average pore area is obtained by a simple variable transform of Equation~(\ref{e:GaussTau}):
\bea
p(\tilde{a}) & = & \left|\frac{{\rm d}\tilde\tau}{{\rm d}\tilde{a}}\right|\,g(\sqrt{\pi/\tilde{a}}) \nonumber \\
& = & \sqrt{\frac{x}{8\,\tilde{a}^3\,\bar\tau}}\,e^{-\frac{x\,(\sqrt\pi - \sqrt{\tilde{a}}\,\bar\tau)^2}{2\,\tilde{a}\,\bar\tau}}\ \ . \label{e:pdfa}
\eea

\begin{figure}\begin{flushright}
{\resizebox{14 cm}{!}{\includegraphics{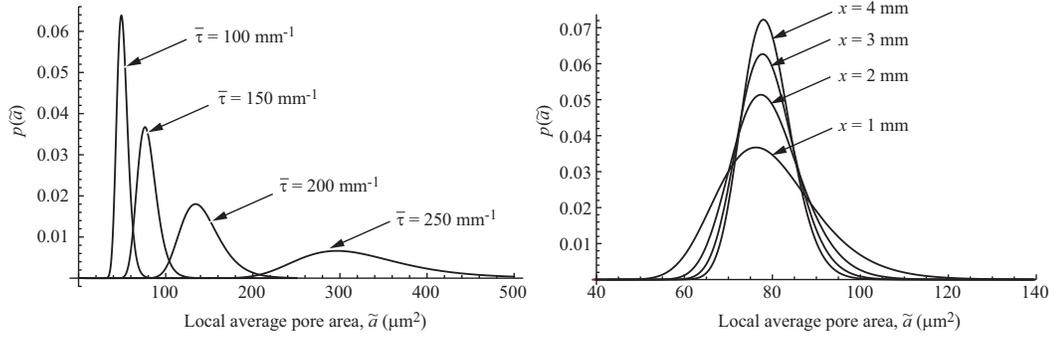}}}
\caption{Probability density function for local average pore
area~$\tilde{a}$ as given by Equation~(\protect\ref{e:pdfa}). Left: influence of process intensity~$\bar\tau$ at scale of inspection~$x = 1~\rm{mm}$; right: influence of scale of inspection~$x$ when process intensity~$\bar\tau = 200~{\rm mm}^{-1}$.} \label{F:pdfa}
\end{flushright}
\end{figure}
The probability density given by Equation~(\ref{e:pdfa}) is plotted for a range of process intensities,~$\bar\tau$ and scales of inspection,~$x$ in Figure~\ref{F:pdfa}. As anticipated, the distribution exhibits a positive skew and narrows with increasing~$\bar\tau$ and~$x$. A consequence of neglecting skewness and using the approximation given by Equation~(\ref{e:approxskew}) for the variance of local average process intensity is that the expected local average pore area obtained as $\bar{a} = \int_0^\infty \tilde{a}\,p(\tilde{a})\,\rmd\tilde{a}$ is slightly greater than~$\pi/\bar\tau^2$ though for~$\bar\tau \ge 200\,{\rm mm}^{-1}$ and $x \ge 1\,{\rm mm}$ the error is less than~2~\%. The variance of local average pore area, is given by
\be\label{e:varaeqn}
\sigma^2(\tilde{a}) = \int_0^\infty (\tilde{a} - \bar{a})^2\,p(\tilde{a})\,\rmd\tilde{a}\ \ .
\ee
\noindent It has not been possible to obtain a closed form solution to this integral though an analytic estimate can be obtained through consideration of Equations~(\ref{e:Nbar}) and~(\ref{e:atilde}). From Equation~(\ref{e:Nbar}) we expect the local average process intensity to be
\be
\widetilde\tau = \frac{\widetilde{N}\,\bar{l}_1}{x}\ .
\ee
\noindent Substituting in Equation~(\ref{e:atilde}), we obtain
\be
\tilde{a} = \frac{\pi\,x^2}{\bar{l}_1^2}\,\frac{1}{{\widetilde{N}}^2}\ \ ,
\ee
\noindent such that
\be
\sigma_x^2(\tilde{a}) = \left(\frac{\pi\,x^2}{\bar{l}_1^2}\right)^2\,\sigma_x^2(1/{\widetilde{N}}^2)\ .
\ee

\begin{figure}
\begin{center}
{\resizebox{8 cm}{!}{\includegraphics{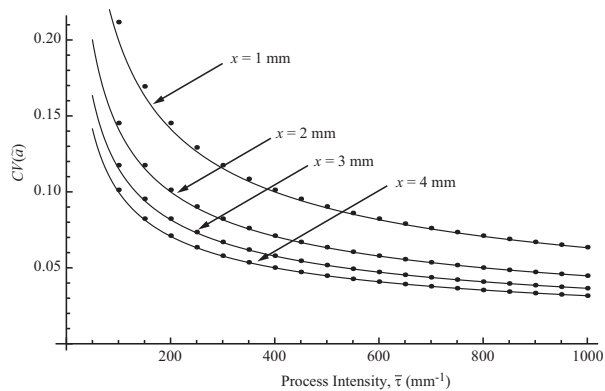}}}
\caption{Coefficient of variation of local average pore
area~$CV_x(\tilde{a})$ plotted against mean process intensity~$\bar\tau$.} \label{F:CVa}\end{center}
\end{figure}

Now, $\widetilde{N}$ is a Poisson variable and since $P(\widetilde{N} = 0) > 0$, $\sigma_x^2(1/{\widetilde{N}}^2)$ is undefined. Typically, we expect~$\bar{N}$ to be sufficiently large that $P(\widetilde{N} = 0)$ is negligible such that $0 < 1/\widetilde{N} \leq 1$. A convenient approximation to the discrete Poisson probability function is the probability density of a Gamma distributed continuous random variable with variance equal to the mean. This probability density is given by
\be
q(\widetilde{N}) = \frac{e^{-\widetilde{N}}\,\widetilde{N}^{\bar{N}-1}}{\Gamma(\bar{N})}\ \ ,
\ee
\noindent such that the probability density of $\nu = 1/{\widetilde{N}}^2$ is given by
\bea
r(\nu)&  = & \left| \frac{\rmd \widetilde{N}}{\rmd \nu}\right|\, q(\nu) \nonumber \\
         & = & \frac{e^{-1/\sqrt{\nu}}\,\nu^{-(1+\bar{N}/2)}}{2\,\Gamma(\bar{N})}
\eea
\noindent The distribution has mean~$\bar\nu = 1/\left((\bar{N}-1)(\bar{N}-2)\right)$ and variance
\bea
\sigma^2(\nu) = \sigma^2(1/{\widetilde{N}}^2) & = & \frac{4\,\bar{N}-10}{(\bar{N}-1)^2(\bar{N}-2)^2(\bar{N}-3)(\bar{N}-4)} \nonumber \\
& \approx & \frac{4}{{\bar{N}}^5} \hspace{2.5 cm} {\rm for\  }\bar{N} \gg 20
\eea
\noindent Such that
\be
\sigma_x^2(\tilde{a}) \approx \frac{4}{{\bar{N}}^5}\,\left(\frac{\pi\,x^2}{\bar{l}_1^2}\right)^2 \ \ .
\ee
\noindent Substituting for~$\bar{N}$ from Equation~(\ref{e:Nbar}) yields
\be
\sigma^2(\tilde{a})  \approx \frac{4\,\pi^2\,\bar{l}_1}{x\,\bar\tau^5}\ \ .
\ee
\noindent Since $\bar{a} = \pi/\bar\tau^2$, it follows that the coefficient of variation of local average pore area is approximated by
\bea
CV(\tilde{a})& \approx & 2\,\sqrt{\frac{\bar{l}_1}{x\,\bar\tau}}\nonumber \\
 & \approx & \frac{2}{\sqrt{x\,\bar\tau}}\ \ .\label{e:CVa_Approx}
\eea
\noindent We observe that the influence of zone size and process intensity is coupled such that the coefficient of variation depends only on the dimensionless product,~$x\,\bar\tau$. The coefficient of variation of local average pore area, as calculated via numerical integration of Equation~(\ref{e:varaeqn}) is plotted against mean process intensity in Figure~\ref{F:CVa}. The solid lines represent the approximation given by Equation~(\ref{e:CVa_Approx}). We note that Schweers and L\"{o}ffler~\cite{schweers} report a coefficient of variation of local flow velocity through a porous nonwoven filter of~0.3 at the~0.5~mm scale, consistent with the values of~$CV(\tilde{a})$ plotted in Figure~\ref{F:CVa}.

\subsection{Distribution of local average pore diameter}
Although the voids generated by random fiber processes are irregular convex polygons, it is often convenient to characterize their dimensions by an equivalent diameter, rather than by area. A good candidate for such a measure is the equivalent diameter determined from the hydraulic radius and defined as the ratio of the area of a polygon to its perimeter. Despite the established utility of this measure, to calculate it for our system we would require knowledge of the joint probability density of polygon perimeter and area, which is unknown. Two alternative measures have been employed previously: the diameter of the largest circle that can be inscribed within a polygon~\cite{miles64,ostoja} and the diameter of a circle with the same area as a polygon~\cite{ds97,cl}. The expected diameter of inscribed circles is given by Equation~(\ref{e:dbarMiles}), and from Equation~(\ref{e:abarMiles}) it follows that the expected diameter of circles with the same area as a polygons is~$\bar{d}_{\rm eq} = 2/\bar\tau$. So, in both cases we see that the expected pore diameter is inversely proportional to the process intensity~$\bar\tau$, and we expect this dependency to hold for the local averages such that~$\tilde{d} = \tilde{d}_{\rm eq}/2 =  1/\widetilde\tau$. Accordingly, we proceed to derive the probability density for the local averages of diameters of inscribed circles and note that this can be readily scaled to give the probability density of equivalent diameters; for convenience, we will term the diameter of an inscribed circle the `pore diameter'.

The probability density of local average pore diameter is obtained by a variable transform of Equation~(\ref{e:GaussTau}):

\bea
q(\tilde{d}) & = & \left|\frac{{\rm d}\tau}{{\rm d}\tilde{d}}\right|\,g(1/\tilde{d}) \\
& = & \frac{1}{{\tilde{d}}^2}\,\sqrt{\frac{x}{2\,\pi\,\bar\tau}}\,e^{-\frac{x\,(1 - \tilde{d}\,\bar\tau)^2}{2\,{\tilde{d}}^2\,\bar\tau}}\ \ . \label{e:pdfd}
\eea

\begin{figure}\begin{flushright}
{\resizebox{14 cm}{!}{\includegraphics{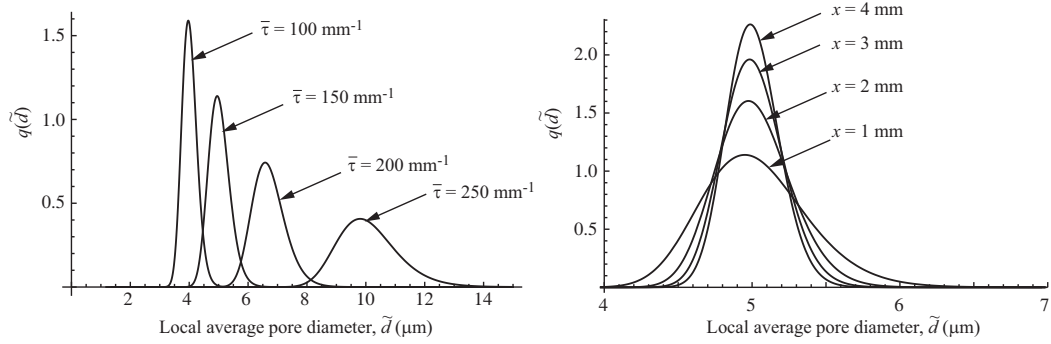}}}
\caption{Probability density function for local average pore
diameter~$\tilde{d}$ as given by Equation~(\protect\ref{e:pdfd}). Left: influence of process intensity~$\bar\tau$ at scale of inspection~$x = 1~\rm{mm}$; right: influence of scale of inspection~$x$ when process intensity~$\bar\tau = 200~{\rm mm}^{-1}$.} \label{F:pdfd}
\end{flushright}
\end{figure}

Following our earlier treatment, we estimate the variance of local average pore diameter.

\bea
\tilde{d} & = & \frac{1}{\widetilde\tau} = \frac{x}{\widetilde{N}\,l_1} \\
\sigma_x^2(\tilde{d}) & = & \frac{x^2}{l_1^2}\,\sigma_x^2(1/\widetilde{N})
\eea
\noindent Again approximating the Poisson distribution for~$\widetilde{N}$ by a gamma distribution with variance equal to the mean, we obtain
\be
\sigma_x^2(1/\widetilde{N}) \approx \frac{1}{{\overline{N}}^3}\ \ ,
\ee
\noindent such that
\bea
\sigma_x^2(\tilde{d}) & \approx & \frac{x^2}{l_1^2\,{\overline{N}}^3} \\
& \approx & \frac{l_1}{x\,\bar\tau^3}
\eea
\noindent and
\bea
CV_x(\tilde{d}) & \approx & \sqrt{\frac{l_1}{x\,\bar\tau}}\\
 & \approx & \frac{1}{\sqrt{x\,\bar\tau}} \label{e:CVd_Approx}
\eea

\begin{figure}
\begin{center}
{\resizebox{8 cm}{!}{\includegraphics{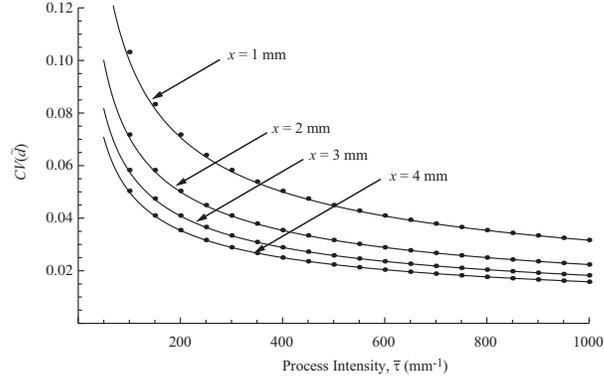}}}
\caption{Coefficient of variation of local average pore
diameter~$CV_x(\tilde{d})$ plotted against mean process intensity~$\bar\tau$.} \label{F:CVd}\end{center}
\end{figure}

Note that our estimate of the coefficient of variation of local average pore diameter given by Equation~(\ref{e:CVd_Approx}) applies for both pore inscribed pore diameters and the diameters of circles with the same area as pores. Comparing Equations~(\ref{e:CVa_Approx}) and~(\ref{e:CVd_Approx}) we observe that our approximation for the coefficient of variation of local average pore area is double that of the local average pore diameter, so again depends only on the dimensionless product~$\bar\tau\,x$. The coefficient of variation of local average pore diameter as calculated via numerical integration to obtain the moments of Equation~(\ref{e:pdfd}) is plotted against mean process intensity in Figure~\ref{F:CVd}. The solid lines represent the approximation given by Equation~(\ref{e:CVd_Approx}).

In the introduction, we noted that the coefficients of variation of polygon perimeter and diameter were~$\sqrt{3}/2$ and approximately~2, respectively, and thus are independent of process intensity. To a first approximation we expect the `diameter' of convex polygons to be proportional to their perimeter, such that the coefficient of variation of diameters is constant also{\footnote{There is good experimental evidence that this is the case~\protect{\cite{cl,bliesn,DS_96_JPPS,S_JMS}}}. Here we have seen that the coefficients of variation of the local averages of pore diameter and pore area are not constant, and both exhibit the same strong dependence on the mean process intensity and the scale of inspection.

\section{Conclusions}
Established theory for the global average polygon dimensions arising from planar Poisson line processes has been extended to give the distributions of local average polygon dimensions. Assuming that the number of lines passing through contiguous square regions of the network has a Poisson distribution, the variance of local average process intensity has been shown to be proportional to the mean process intensity and inversely proportional to the side of the square regions. Through an approximation, the skewness of the distribution has been shown to be inversely proportional to the square root of the expected number of lines passing through each region; this result has been confirmed by Monte Carlo simulation.

For typical values of process intensity observed in heterogeneous fibrous materials, the skewness of the distribution of local process intensity rapidly approaches zero and the distribution can be assumed to be Gaussian. On this basis, the distribution of local average polygon area and that of local average pore diameter have been derived, along with approximate expressions for their coefficients of variation. We find that the coefficient of variation of pore diameter is inversely proportional to the square root of the product of process intensity and zone size; the coefficient of variation of pore area is double that of pore diameter.


\section*{References}
\begin{thebibliography}{99}
\bibitem{cl} H.\ Corte and E.H.\ Lloyd. Fluid flow through paper
and sheet structure. In {\bf Consolidation of the Paper Web} {\it
Trans.\ IIIrd Fund.\ Res.\ Symp.} (F.\ Bolam, ed.), pp981-1009,
BPBMA, London, 1966.

\bibitem{ds96} C.T.J.\ Dodson and W.W.\ Sampson. The effect of paper
formation and grammage on its pore size distribution. {\it J.\
Pulp Pap.\ Sci.} {\bf 22}(5):J165-J169, 1996.

\bibitem{ad85} M.S.\ Abdel-Ghani and G.A.\ Davies. Simulation of
non-woven fibre mats and the application to coalescers. {\it
Chem.\ Eng.\ Sci.} {\bf 40}(1):117-129, 1985.

\bibitem{pk} H.W.\ Piekaar and L.A.\ Clarenburg. Aerosol filters---Pore size
distribution in fibrous filters. {\it Chem.\ Eng.\ Sci.} {\bf
22}():1399-1408, 1967.

\bibitem{ostoja} J.\ Castro and M.\ Ostoja-Starzewski. Particle seiving in a random fibre network. {\it Appl. Math.\ Model.} {\bf 24}(8-9):523-534, 2000.

\bibitem{endo} M.\ Endo, H.\ Muramatsu, T.\ Hayashi, Y.A.\ Kim, M.\ Terrones and M.S.\ Dresselhaus. 'Buckypaper' from coaxial nanotubes. {\it Nature} {\bf 433}(7025):476, 2005.

\bibitem{hall} L.J.\ Hall, V.R.\ Coluci, D.S.\ Galv\~{a}o, M.K.\ Kozlov, M.\ Zhang, S.O.\ Dantes and R.H.\ Baughman. Sign change of Poisson's ratio for carbon nanotube sheets. {\it Science} {\bf 320}(5875):504-507, 2008.

\bibitem{paul} D.R.\ Paul and L.M.\ Robeson. Polymer nanotechnology: nanocompsoites. {\it Polymer} {\bf 49}(15):3187-3204, 2008.


\bibitem{eichhornreview} S.J.\ Eichhorn {\it et al.} Review: current international research into cellulose nanofibres and nanocomposites. {\it J.\ Mater.\ Sci.} {\bf 45}(1):1-33, 2010.

\bibitem{pham} Q.P.\ Pham, U.\ Sharma and A.G. Mikos. Electrospinning of polymeric nanofibers for tissue engineering applications: a review. {\it Tissue Eng.} {\bf 12}(5):1197-1211, 2006.

\bibitem{sill} T.J.\ Sill, H.A.\ von Recum. Electro spinning: Applications in drug delivery and tissue engineering. {\it Biomater.} {\bf 29}(13):1989-2006, 2008.


\bibitem{miles64} R.E.\ Miles. Random polygons determined by random lines in
a plane. {\it Proc.\ Nat.\ Acad.\ Sci.\ USA} 52
901-907,1157-1160 (1964).

\bibitem{tanner} J.C.\ Tanner.
The proportion of quadrilaterals formed by random lines in a
plane. {\em J.\ Appl.\ Probab.} 20(2),  400-404 (1983).

\bibitem{crainmiles} I.K.\ Crain and R.E.\ Miles. Monte Carlo estimates of the distributions
of the random polygons determined by random lines in the plane.
{\it J.\ Statist.\ Comput.\ Simul.} {\bf 4}:293-325, 1976.


\bibitem{george} E.I.\ George. Sampling random polygons. {\it J.\ Appl.\ Prob.} {\bf 24}(3):557-573, 1987.

\bibitem{PlinProp} C.T.J.\ Dodson and W.W.\ Sampson. Planar line processes for
void and density statistics in thin stochastic fibre networks.
{\it  J.\ Statist.\ Phys.} {\bf 129}(2):311-322, 2007.


\bibitem{kalcort1} O.\ Kallmes and H.\ Corte. The structure of paper, I. The
statistical geometry of an ideal two dimensional fiber network.
{\it Tappi J.} {\bf 43}(9):737-752, 1960. {\it Errata:} {\bf
44}(6):448, 1961.

\bibitem{EichSamp2} S.J.\ Eichhorn and W.W.\ Sampson. Relationships between specific surface area and pore size in electrospun polymer fibre networks. {\it J.\ Roy.\ Soc.\ Interface} {\bf 7}(45):641-649, 2010.

\bibitem{berhan} L.\ Berhan, Y.B.\ Yi and A.M.\ Sastry. Effect of nanorope waviness on the effective moduli of nanotube
sheets. {\it J.\ Appl.\ Phys.} {\bf 95}(9): 5027–5034, 2004.

\bibitem{ds97} C.T.J.\ Dodson and W.W. Sampson. Modeling a class of
stochastic porous media. {\em App.\ Math.\ Lett.} 10(2),
87-89 (1997).

\bibitem{mppsd} W.W.\ Sampson. A multiplanar model for the pore radius
distribution in isotropic near-planar stochastic fibre networks.
{\it J. Mater.\ Sci.} {\bf 38}(8):1617-1622, 2003.

\bibitem{chatterjee} A.P.\ Chatterjee. Nonuniform fiber networks and fiber-based composites: Pore size distributions and elastic moduli. {\it J.\ Appl.\ Phys.} {\bf 108}: 063513, 2010.

\bibitem{wwsAppita} C.T.J.\ Dodson, A.G.\ Handley, Y.\ Oba and W.W.\ Sampson.
The pore radius distribution in paper. Part~I: The effect of
formation and grammage. {\it Appita J.} {\bf 56}(4):275-280, 2003.

\bibitem{JFBloch} J.-F.\ Bloch and S.\ Rolland du Roscoat. Three-dimensional structural analysis. In {\bf Advances in Pulp and Paper Research, Oxford 2009}. (ed. S.J.\ I'Anson), Proc.\ 14th Fund.\ Res.\ Symp.\}, pp599-664. PPFRS, Manchester, 2009.

\bibitem{jaganathan} S.\ Jaganathan, H.V.\ Tafreshi and B.\ Pourdeyhimi. Modeling liquid porosimetry in modeled and imaged 3-D fibrous microstructures. {\it J.\ Coll.\ Interf.\ Sci.} {\bf326}(1):166-175, 2008.

\bibitem{tomba} E.\ Tomba, P.\ Facco, M.\ Roso, M.\ Modesti, F.\ Bezzo and M.\ Barolo. Artificial vision system for the automatic measurement of interfiber pore characteristics and fiber diamater in nanofiber assemblies. {\it Ind.\ Eng.\ Chem.\  Res.} {\bf 49}(6):2957-2968, 2010.

\bibitem{ctjdJRSS} C.T.J.\ Dodson. Spatial variability and the theory of sampling in random
fibrous networks. {\it J.\ Roy.\ Statist.\ Soc.} {\bf B 33}(1):88-94, 1971.

\bibitem{deng} M.\ Deng and C.T.J.\ Dodson. {\bf Paper: An Engineered
Stochastic Structure}.
Tappi Press, Atlanta, 1994.


\bibitem{uesaka} D.T.\ Hristopulos and T.\ Uesaka. Structural disorder effects
on the tensile strength distribution of heterogeneous brittle materials with emphasis
on fiber networks. {\it Phys.\ Rev.\ B.} {\bf 70}(6):064108, 2004.

\bibitem{wathen} R.\ Wath\'{e}n and K.\ Niskanen. Strength distributions of running paper webs. {\it J.\ Pulp Paper Sci.} {\bf 32}(3):137-144, 2006.

\bibitem{schweers} E.\ Schweers and F.\ L\"{o}ffler. Realistic modelling of the behaviour of fibrous filters through consideration of filter structure. {\it Powder Tech.} {\bf 80}(3):191-206, 1994.

\bibitem{huang} C.\ Huang, K.\ Willeke, Y.\ Qian, S.\ Grinshpun and V.\ Ulevicius. Method for measuring the spatial variability of aerosol penetration through respirator filters. {\it Am.\ Ind.\ Hyg.\ Assoc.\ J.} {\bf 59}(7):461-465, 1998.

\bibitem{chatterjee2} A.P.\ Chatterjee. A simple model for chracterizing non-uniform fibre-based composites and networks. {\it J.\ Phys.: Cond.\ Matter} {\bf 23}(15):155014, 2011.


\bibitem{holzmeister} A.\ Holzmeister, M.\ Rudisile, A.\ Greiner and J.H.\ Wendorff. Structurally and chemically heterogenous nanofibrous nonwovens via electrospinning. {\it Eur.\ Polym.\ J.} {\bf 43}(12):4859-4867, 2007.
    
    
\bibitem{coleman} R.\ Coleman. Random paths through convex bodies. {\it J.\ Appl.\ Prob.} {\bf 6}(2):430-441, 1969.

\bibitem{CRC_SMTF} W.C.\ Rinaman, C.\ Heil, M.T.\ Strauss, M.\ Mascagni and M.\ Souza. Probability and Statistics. Chapter~7 in {\bf CRC Standard Mathematical Tables and Formulae} (D.\ Zwillinger, ed.), 30th edition. CRC Press, Boca Raton, 1996.

\bibitem{mma} Wolfram Research, Inc., Mathematica, Version 8.0, Champaign, IL, 2010.

\bibitem{Azzalini} A.\ Azzalini. A class of distributions which includes the normal ones. {\it Scand. J.\ Statist.} {\bf 12}(2):171-178, 1985.

\bibitem{bliesn} W.C.\ Bliesner. A study of the porous structure of
fibrous sheets using permeability techniques. {\it Tappi J.} {\bf 47}(7):392-400, 1964.

\bibitem{DS_96_JPPS}  C.T.J.\ Dodson and W.W.\ Sampson. The effect of paper
formation and grammage on its pore size distribution. {\it J.\
Pulp Pap.\ Sci.} {\bf 22}(5):J165-J169, 1996.


\bibitem{S_JMS} W.W.\ Sampson. Comments on the pore radius
distribution in near-planar stochastic fibre networks. {\it J.\
Mater.\ Sci.} {\bf 36}(21):5131-5135, 2001.
\end{thebibliography}
\end{document}